\documentclass[RNAAS]{aastex62}

\begin{document}

\title{WASP-South detection of HD\,219666b transits provides an accurate ephemeris}

\correspondingauthor{C. Hellier}
\email{c.hellier@keele.ac.uk}
\author{C. Hellier}
\affiliation{Astrophysics Group, Keele University, Staffordshire, ST5 5BG, UK}
\author{D. R. Anderson}
\affiliation{Astrophysics Group, Keele University, Staffordshire, ST5 5BG, UK}
\affiliation{Department of Physics, University of Warwick, Gibbet Hill Road, Coventry CV4 7AL, UK}
\author{S. Gill}
\affiliation{Department of Physics, University of Warwick, Gibbet Hill Road, Coventry CV4 7AL, UK}
\author{R. G. West}
\affiliation{Department of Physics, University of Warwick, Gibbet Hill Road, Coventry CV4 7AL, UK}


\section{} 
The hot-Neptune HD\,219666b (= TOI-118.01; \citealt{2019A&A...623A.165E}) was an early discovery from Sector 1 of the {\it TESS\/} transit survey \citep{2016SPIE.9904E..2BR}. Being a rare ``Neptune desert'' planet, and transiting a bright, $V$ = 9.9 star, it is a prime target for further study including atmospheric characterisation. Targeting exoplanet transits with major facilities such as {\it HST\/} and the imminent {\it JWST\/} depends on an accurate ephemeris.  The transit period presented by Esposito et\,al (2019) derives primarily from 4 transits spanning 18 days and has an uncertainty that will amount to several hours by the time that {\it TESS\/} re-observes this region of sky, likely to be in 2021.

Here we report a recovery of the transit in WASP-South data, enabling an ephemeris with a period uncertainty that is a factor of 60 smaller.   The 8-camera, WASP-South transit survey \citep{2006PASP..118.1407P} has produced 150 transiting exoplanets (e.g.~\citealt{2019MNRAS.482.1379H}). The transit depth of HD\,219666b, at only  0.17\%, is shallower than that of any WASP planet (for example the bright-star, super-Neptune WASP-166b has a transit depth of 0.28\%; \citealt{2019MNRAS.488.3067H}). Nevertheless, at $V$ = 9.9, and being relatively isolated on the sky, HD\,219666 was in the ``sweet spot'' for WASP-South, both when using 200-mm lenses (observing HD\,219666 from 2010--2012), and when using 85-mm lenses (observing 2012--2014). A total of 148\,000 photometric data points were obtained on this star. 

The standard WASP transit-search code \citep{2007MNRAS.380.1230C} successfully finds the transit, from coverage of 24 partial or full transits over the period 2010 July to 2014 Oct. The code reports the dip as 0.2\%\ deep and 2-hr wide, values consistent with those from {\it TESS} (0.17\%\ deep and 2.5-hr wide).  The resulting transit ephemeris is:

\[ {\rm WASP\ (JD\ TDB)} = (245\,5788.6920 \pm 0.0043) + N \times (6.03446 \pm 0.00007) \] 
which compares with the \citet{2019A&A...623A.165E} ephemeris:
\[ {\rm TESS\ (JD\ TDB)} = (245\,8329.1996 \pm 0.0012) + N \times (6.03607 \pm 0.00064). \] 

The {\it TESS}/WASP period ratio is 1.00027 and thus the chances of a period match to this accuracy being spurious are 1-in-2000, though bear in mind that we have also been looking for period matches with several hundred other {\it TESS\/} candidates where a WASP match is feasible.     

At the epoch of the {\it TESS\/} observation the WASP ephemeris predicts the transit to an accuracy of 0.03 day, or 0.005 in phase.  Thus the chances of a spurious match within this window are 1-in-100.   The {\it TESS\/} epoch actually locates at 420.99999 cycles on the WASP ephemeris.  Thus the combination of the period match and the phase match give confidence that the WASP detection is real, with a likelihood of being spurious below 1\%. 

Taking the difference between the epochs as exactly 421 cycles then leads to the more accurate ephemeris:

\[ {\rm Combined\ (JD\ TDB)} = (245\,8329.1996 \pm 0.0012) + N \times (6.034460 \pm 0.000011). \] 

The combined ephemeris has a drift of only one minute per year, and should prove useful to those scheduling observations of HD\,219666b. The early WASP timing can also contribute to long-term monitoring of any period changes. The revised period is 2.5 {\it TESS\/} error bars from the {\it TESS\/} period, where we note that the last of the 4 {\it TESS\/} transits was at a time of excess noise in the {\it TESS\/} data, which may have affected that timing.    

\begin{figure}[t]
\begin{center}
\includegraphics[scale=0.85,angle=0]{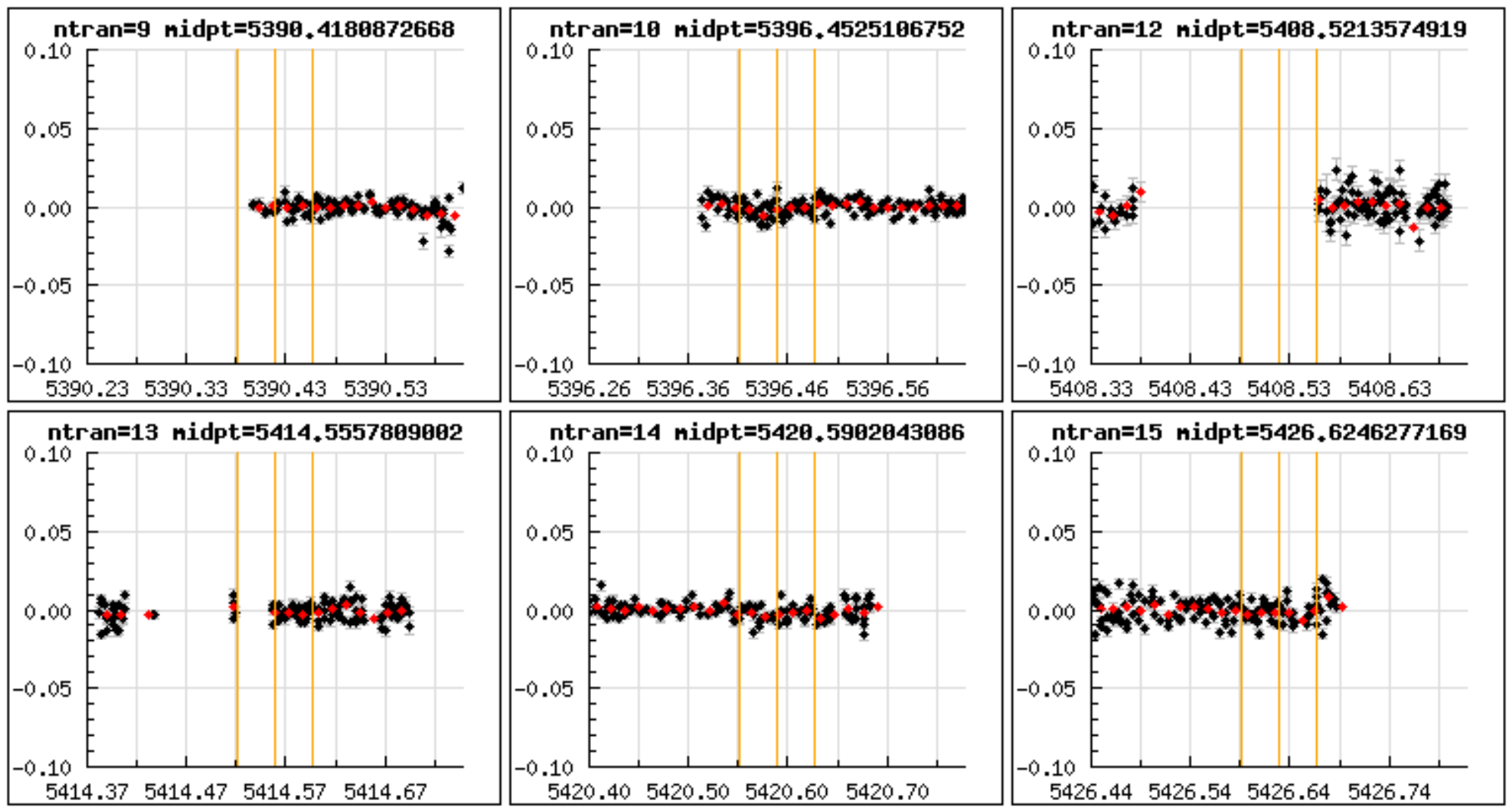}
\caption{Portions of the 200-mm WASP-South data on HD\,219666. The orange lines mark transit times as found by the search algorithm. There are likely detections of the shallow, 0.17\%-deep transit at midpoint times: 245\,5396.45, 245\,5420.59 \&\ 245\,5426.62, and these will be the earliest transits of HD\,219666b recorded.  While the individual transit events are not necessarily convincing to the eye, the match in both period and phase of the search-algorithm output gives confidence that the detection is real.}
\end{center}
\end{figure}

\bibliographystyle{mnras}
\bibliography{biblio}

%
%
%
%
%
%
%

\end{document}